\documentclass[fleqn,usenatbib]{mnras}

\usepackage[dvips]{epsfig}
\usepackage{float}

\usepackage{newtxtext,newtxmath}

\usepackage[T1]{fontenc}
\usepackage{ae,aecompl,pdflscape}

\usepackage{amsmath}	
\usepackage{amssymb}
\usepackage[colorinlistoftodos]{todonotes} 
\usepackage{color, soul}	
\usepackage{xcolor} 

\def\ha{H$\alpha$}

\def\kms {$\rm km\,s^{-1}$}

\title[Precessing winds from SMBHs?]{Precessing winds from the nucleus of the prototype Red Geyser?}

\author[R. A. Riffel et al.]{Rogemar A. Riffel,$^{1,2}$\thanks{E-mail: rogemar@ufsm.br (RAR)}
Rodrigo S. Nemmen,$^{3}$
Gabriele S. Ilha,$^{1,2}$
Sandro B. Rembold,$^{1,2}$
\newauthor Namrata Roy,$^{4}$
Thaisa Storchi-Bergmann,$^{5,2}$
Rogerio Riffel,$^{5,2}$
Kevin A. Bundy,$^{6}$
\newauthor Alice D. Machado,$^{1,2}$
Nicolas D. Mallman,$^{5,2}$
Jaderson S. Schimoia,$^{7,2,1}$
\newauthor Luiz N. da Costa $^{2,8}$
and Marcio A. G. Maia$^{2,8}$
\\
$^{1}$Universidade Federal de Santa Maria, CCNE, Departamento de F\'\i sica,  
 97105-900, Santa Maria, RS, Brazil\\
$^{2}$Laborat\'orio Interinstitucional de e-Astronomia - LIneA, Rua General Jos\'e Cristino 77, Rio de Janeiro, RJ - 20921-400, Brazil\\
$^{3}$Universidade de S\~ao Paulo, IAG, Departamento de Astronomia, S\~ao Paulo, SP 05508-090, Brazil\\
$^{4}$Department of Astronomy and Astrophysics, University of California, 1156 High Street, Santa Cruz, CA 95064\\
$^{5}$Universidade Federal do Rio Grande do Sul, IF, CP 15051, Porto Alegre 91501-970, RS, Brazil\\
$^{6}$UCO/Lick Observatory, University of California, Santa Cruz, 1156 High St., Santa Cruz, CA 95064, USA\\
$^{7}$Universidade Federal de Santa Catarina, Physics Department, 88036-000, Florian\'opolis, SC, Brazil\\
$^{8}$Observat\'orio Nacional - MCT, Rua General Jos\'e Cristino 77, Rio de Janeiro, RJ -- 20921-400, Brazil
}

\date{Accepted XXX. Received YYY; in original form ZZZ}

\pubyear{2019}

\begin{document}
\label{firstpage}
\pagerange{\pageref{firstpage}--\pageref{lastpage}}
\maketitle

\begin{abstract}
Super-massive black holes (SMBH)  are present at the center of most galaxies, with the related mass accretion processes giving origin to outflows in Active Galactic Nuclei (AGN). It has been presumed that only intense winds from luminous AGN were able to suppress star formation until the discovery of a new class of galaxies with no recent star formation and with the nucleus in a quiescent state showing kpc scale outflows. 
We used SDSS MaNGA and Gemini Integral Field Spectroscopy of the prototype Red Geyser Akira and found that the orientation of the outflow changes by about 50$^\circ$ from its nucleus to kpc scales. A possible interpretation is that the outflow is produced by a precessing accretion disk due to a misalignment between the orientation of the disk and the spin of the SMBH.  The precession of the central source is also supported by a similar change in the orientation of the ionization pattern. Although similar behavior has commonly being reported for collimated relativistic jets, the precession of an AGN wide wind is reported here for the first time, implying on a larger work surface of the wind, which  in turn increases the star formation suppression efficiency of the outflow.
\end{abstract}




\section{Introduction} 

Most galaxies containing a spheroidal component (elliptical galaxies and bulges of spiral galaxies) seem to host a central super-massive black hole \citep[SMBH; ][for a review]{heckman14} that interacts with its host galaxy via capture of matter and  emission of radiation, particles and winds from its surroundings. These processes are witnessed in Active Galactic Nuclei (AGN), and seem to play an important role in the evolution of the host galaxies. Cosmological simulations that do not consider the presence of SMBHs and their mechanisms of feeding and feedback that are present in the AGN phase, result in galaxies that are much more massive than those observed \citep{benson03,dimateo05,springel05,bower06}. 
Massive outflows triggered as a consequence of the accretion process to the SMBH can regulate and couple the growth of the galactic bulge and the SMBH \citep[e.g., ][]{hopkins05, cattaneo09}, explaining the relation between the mass of the SMBH and stellar velocity dispersion of the bulge. 

The narrow line regions (NLR) are the signposts of the central activity of SMBHs, since they are generally expected to present a bi-symmetric emission pattern in ionized gas
\citep[e.g.,][]{antonucci93,urry95,harrison18}.  Puzzling enough, Hubble Space Telescope (HST) narrow-band [O\,{\sc iii}]$\lambda$5007 images of a sample of 60 nearby Seyfert galaxies show that the bi-conical shape of the NLR is not as common as expected \citep{schmitt03} and gas outflows are seen only in 33\% of Seyfert galaxies, as revealed by long-slit spectroscopy of 48 nearby AGN \citep{fischer13}. On the other hand, ionized gas outflows in the inner kpc have been commonly observed using optical \citep[e.g.,][]{sm14,lena15,cresci15,karouzos16a,karouzos16b,venturi18} and near-infrared \citep[e.g.,][]{riffel14,barbosa14,rogemar18} integral field spectroscopy (IFS) of nearby active galaxies.
Another puzzling result is that outflows seem to be almost absent for low-luminosity AGN, showing increasing extent and power as the AGN luminosity increases \citep[e.g.][]{ilha18}.

Recently, an intriguing result was reported on a sample of galaxies with no recent star formation activity, most of them harboring a very low-luminosity AGN (LLAGN): a bi-polar outflow seen in ionized gas and interpreted as being originated by centrally driven winds due to a radiatively inefficient accretion flow \citep[RIAF; ][]{yuan15} onto the SMBH \citep{cheung16}. This study is based on large-scale (several kpc) data cubes obtained by the Sloan Digital Sky Survey 4th phase (SDSS-IV) Mapping Nearby Galaxies at Apache Point Observatory 
(MaNGA) survey and suggests that such galaxies -- dubbed  'Red Geysers' -- are very common, accounting for 5--10\,\% of the population of galaxies with stellar masses of $M_\star\sim5\times10^{10}$\,M$_\odot$ that do not show recent star formation episodes \citep{cheung16,roy18}. These large scale outflows are capable of suppressing the star formation in this population of galaxies.  
 \citet{penny18} reported the detection of analogous galaxies to Red Geysers in low-mass galaxies ($M_\star\leq5\times10^{9}$\,M$_\odot$). 
However, the angular resolution of the observations ($\sim3^{\prime\prime}$) does not allow to constrain the kinematics in the nuclear region  and therefore the mechanism which produces the outflows in these galaxies is poorly known.

In order to better constrain the gas kinematics,  probing scales 3 times smaller than those probed by MaNGA, 
we obtained IFS with Gemini GMOS of the inner 1.7\,kpc$\times$2.5\,kpc of $mangaid$\,1-217022 -- dubbed as Akira --  to complement the 
information from MaNGA IFU data. 
Akira is the prototype of the Red Geyser class, harbors a LLAGN and presents a well defined large scale  bipolar outflow  \citep{cheung16}. It has a redshift $z$=0.0245, for which 1$^{\prime\prime}$ corresponds to $\sim$500\,pc at the distance of the galaxy, adopting a Hubble constant of $H_0$=71\,\kms\,Mpc$^{-1}$. 

In this letter we report that for Akira the orientation of outflow changes with the distance from the nucleus, with a possible cause being possibly due to the precession of the accretion disk which creates a precessing wind.  
This paper is organized as follows: section 2 describes the data used in this work, while in Sec. 3 we present our results, which are discussed in Sec. 4. Finally, Sec. 5 presents the final remarks.

\section{Data and Measurements}

\subsection{MaNGA data}

The MaNGA \citep{bundy15} fiber-bundle IFU survey  is part of the SDSS-IV \citep{gunn06,blanton17}, 
uses the Baryon Oscillation Spectroscopic Survey \citep[BOSS,][]{smee13}
spectrograph, covering the spectral range from 3622 to 10354\,\AA\ at a resolving power $R\sim$2000 and will obtain IFU observations of $\sim$10,000 nearby galaxies spanning all environments and the stellar mass range $10^9-^{11}$\,M$_\odot$ at a spatial resolution of 1--2 kpc \citep{bundy15,drory15,law15,yan16b,wake17}.
The data cube for Akira is included in the 14th Data Release \citep[DR14,][]{albareti17} of the SDSS. 

The data processing was performed using the MaNGA Data Reduction Pipeline \citep{law16} and the relative flux calibration of the final spectra is better than 5 per cent \citep{yan16a}. Akira was observed with a fiber-bundle of 61 fibers with a diameter of 9 fibers, resulting in a bundle diameter of 22\farcs5 and the reconstructed FWHM is $\sim$2\farcs7.

\begin{figure}
  \centering
\includegraphics[width=0.5\textwidth]{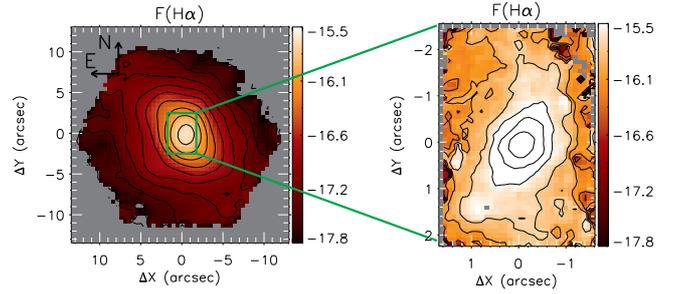} 
  \caption{ \ha\ emission-line flux map using MaNGA (left) and GMOS (right) data. The color bars show the fluxes in logarithmic units of erg\,s$^{-1}$\,cm$^{-2}$\,arcsec$^2$. }
  \label{flux}
\end{figure}

\begin{figure*}
  \begin{tabular}{rlrl}
\includegraphics[width=0.22\textwidth]{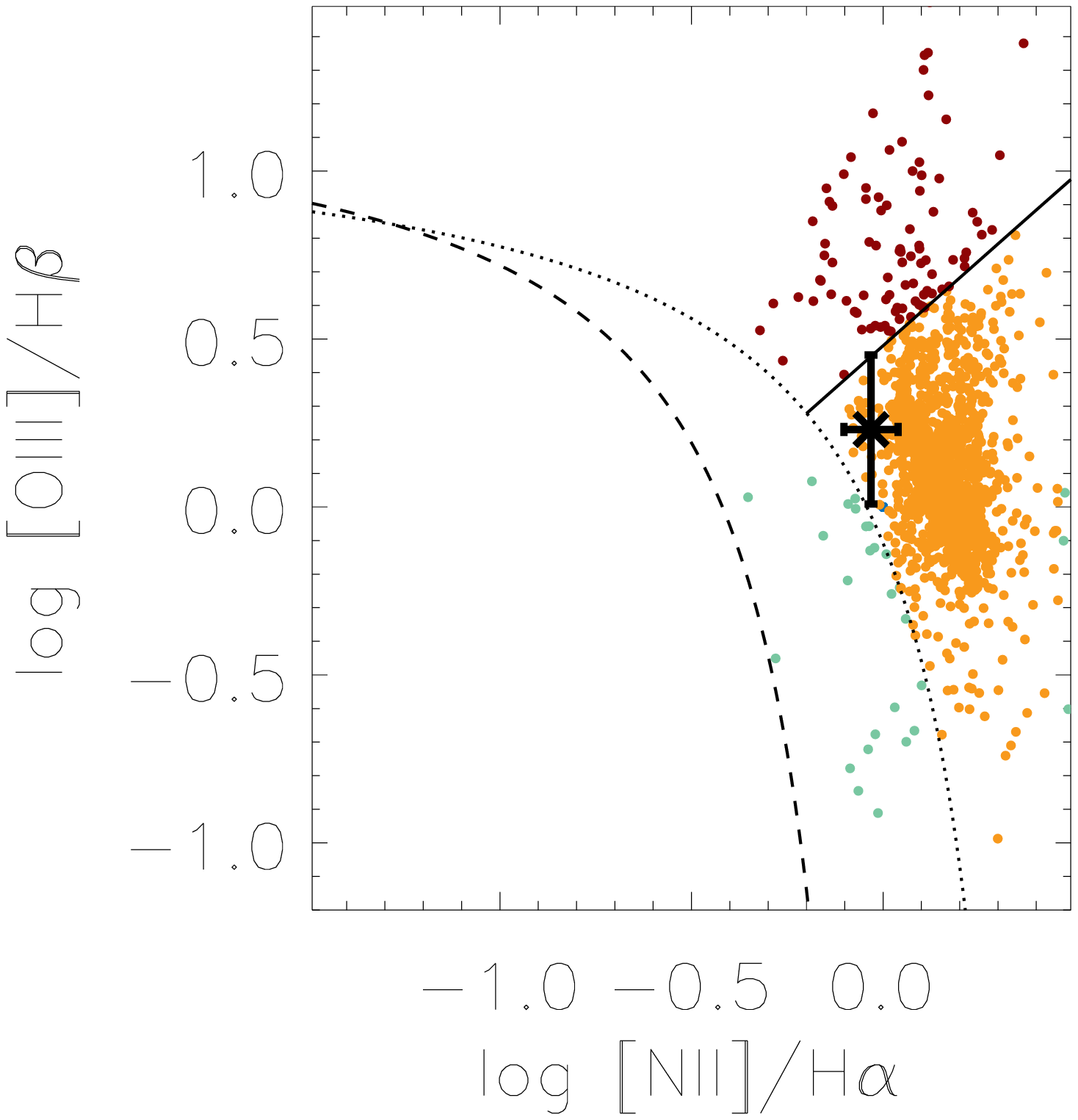} &
\includegraphics[width=0.25\textwidth]{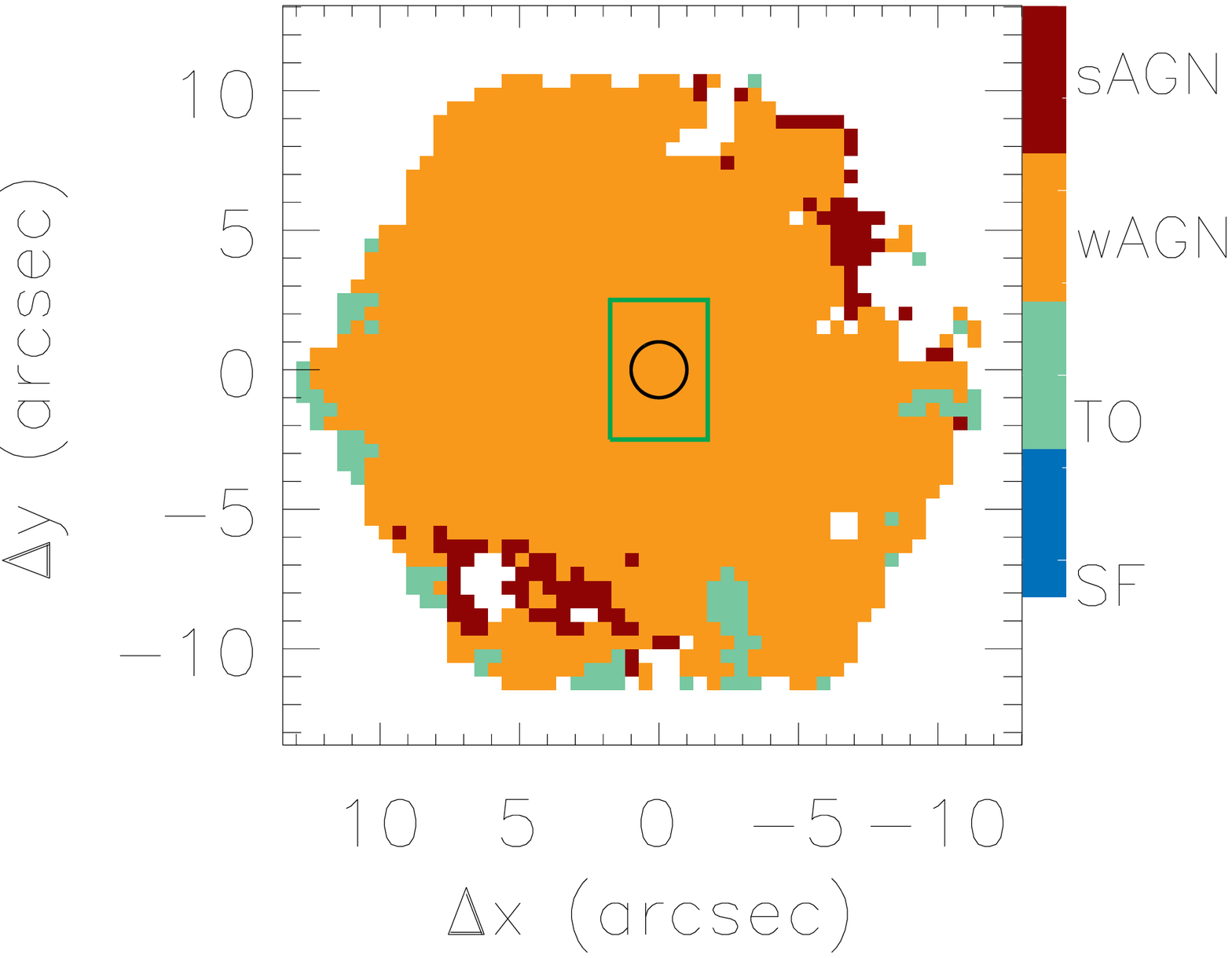} &
\includegraphics[width=0.22\textwidth]{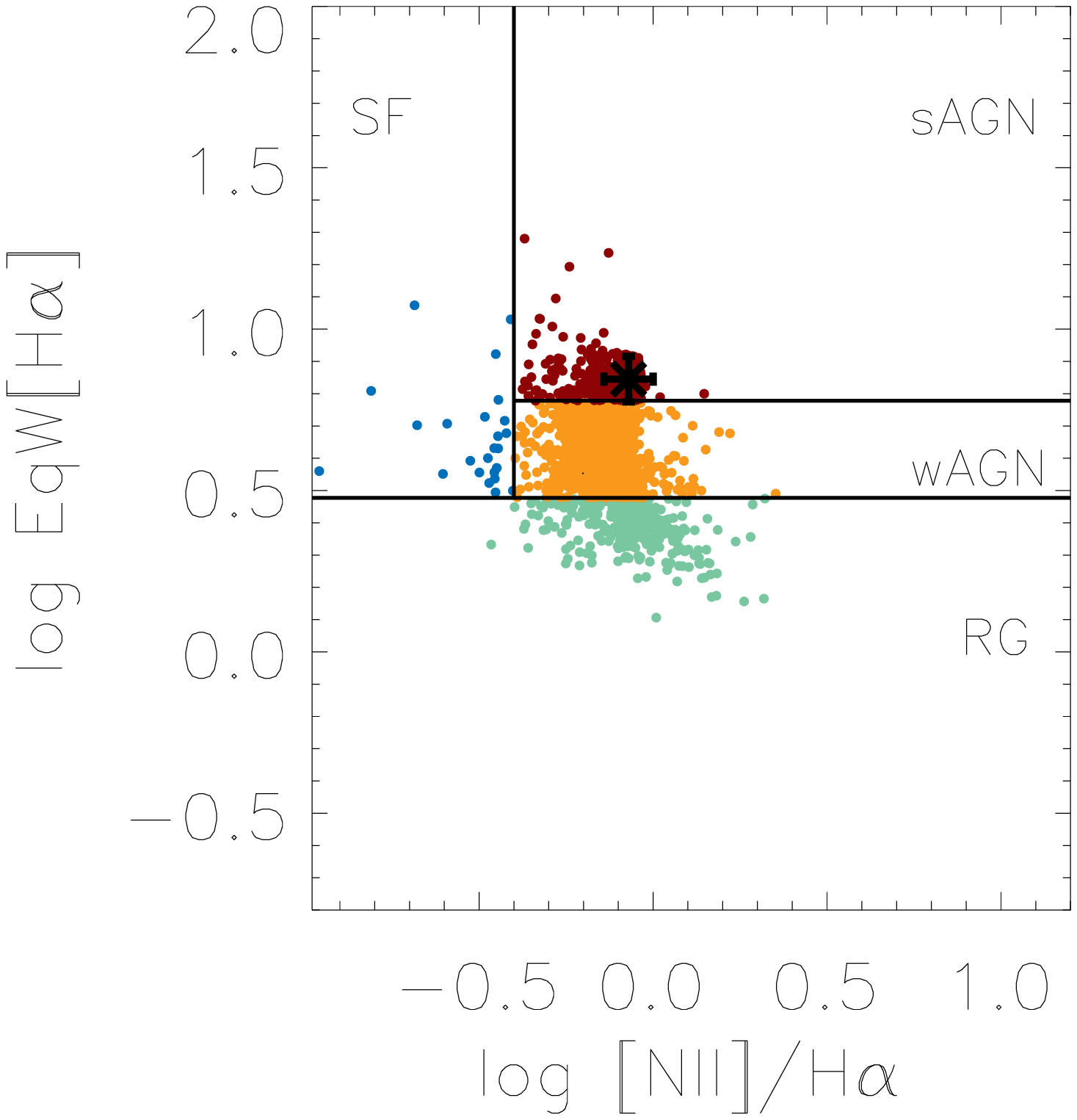} &
\includegraphics[width=0.25\textwidth]{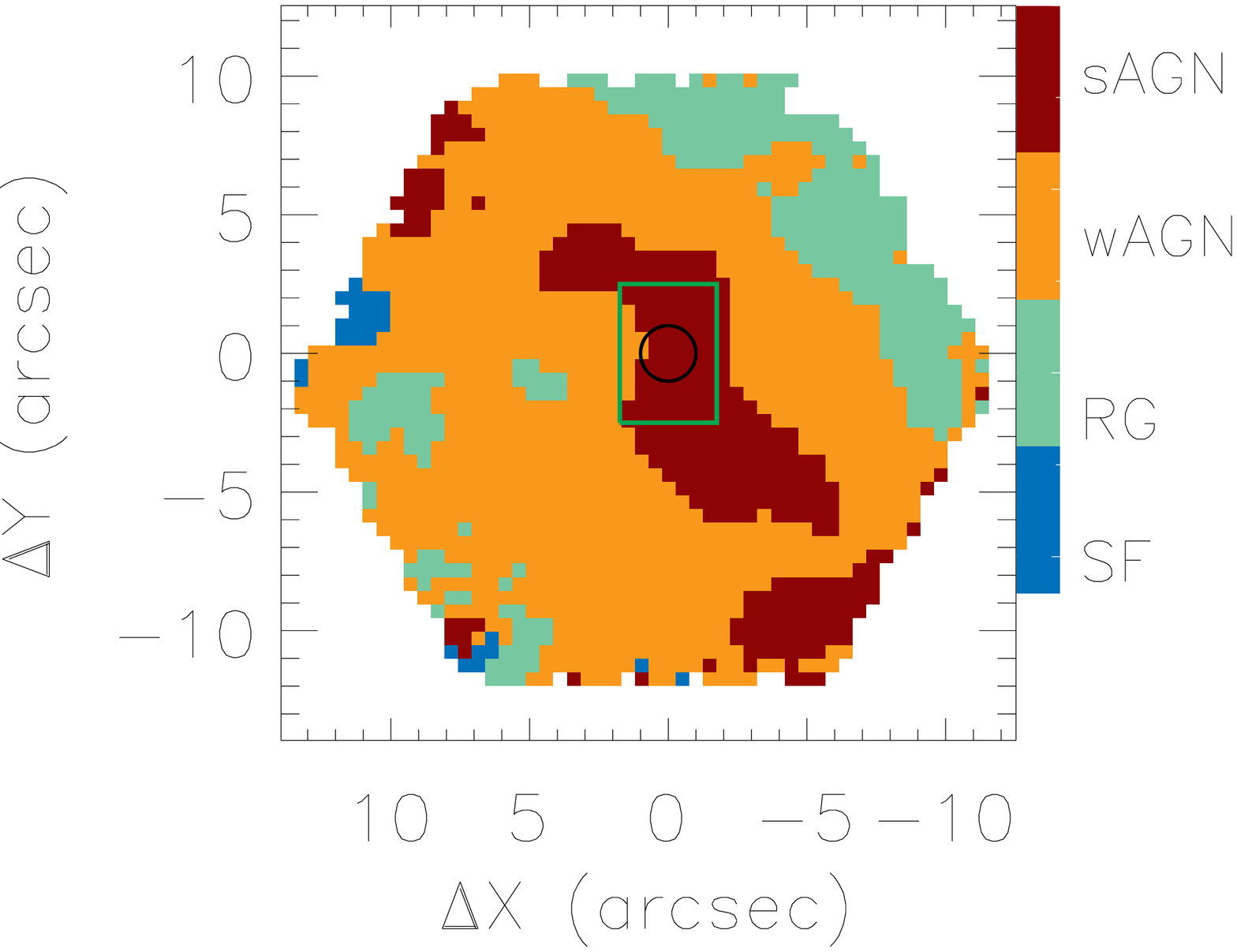} \\

\includegraphics[width=0.22\textwidth]{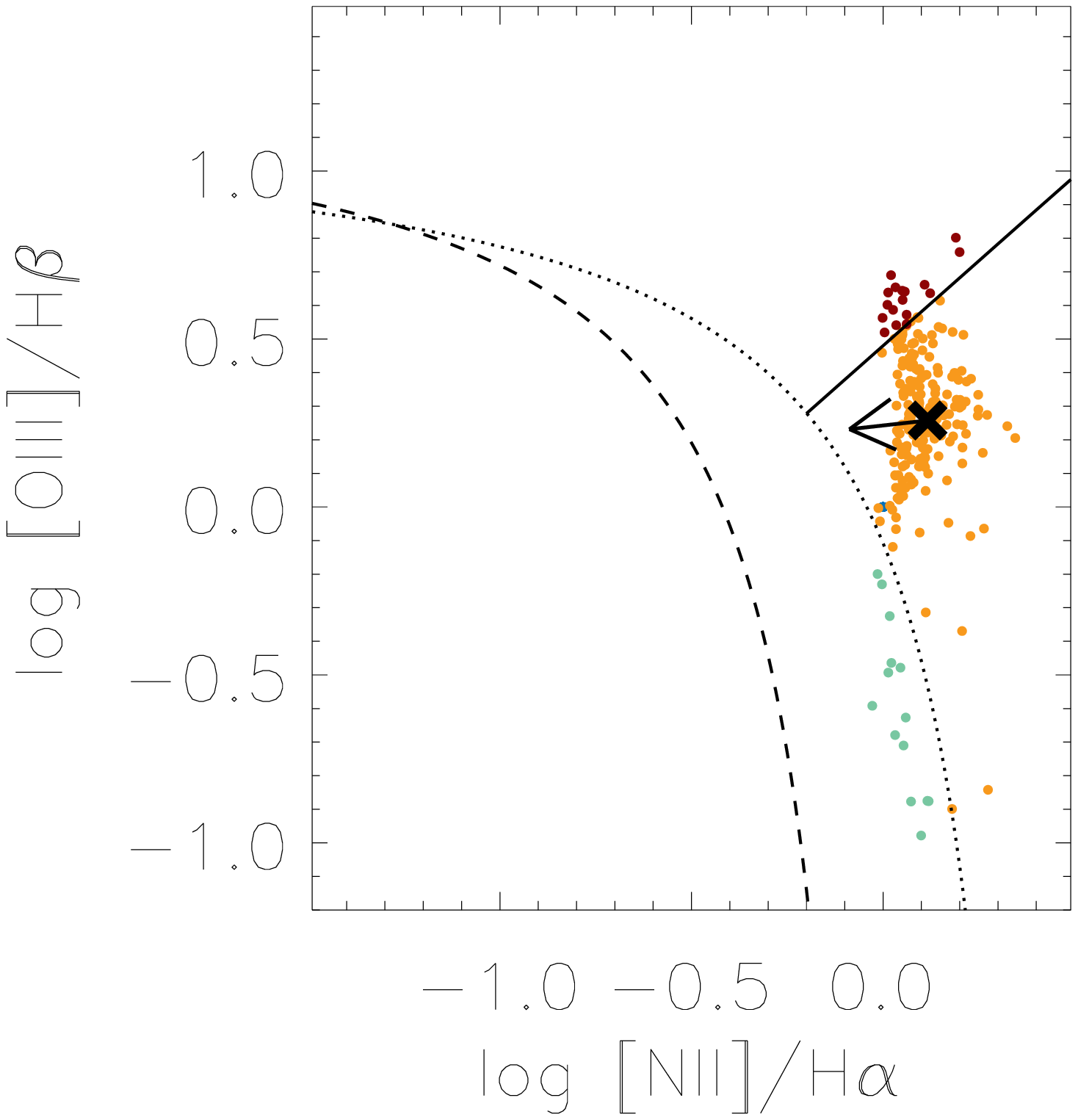} &
\includegraphics[width=0.20\textwidth]{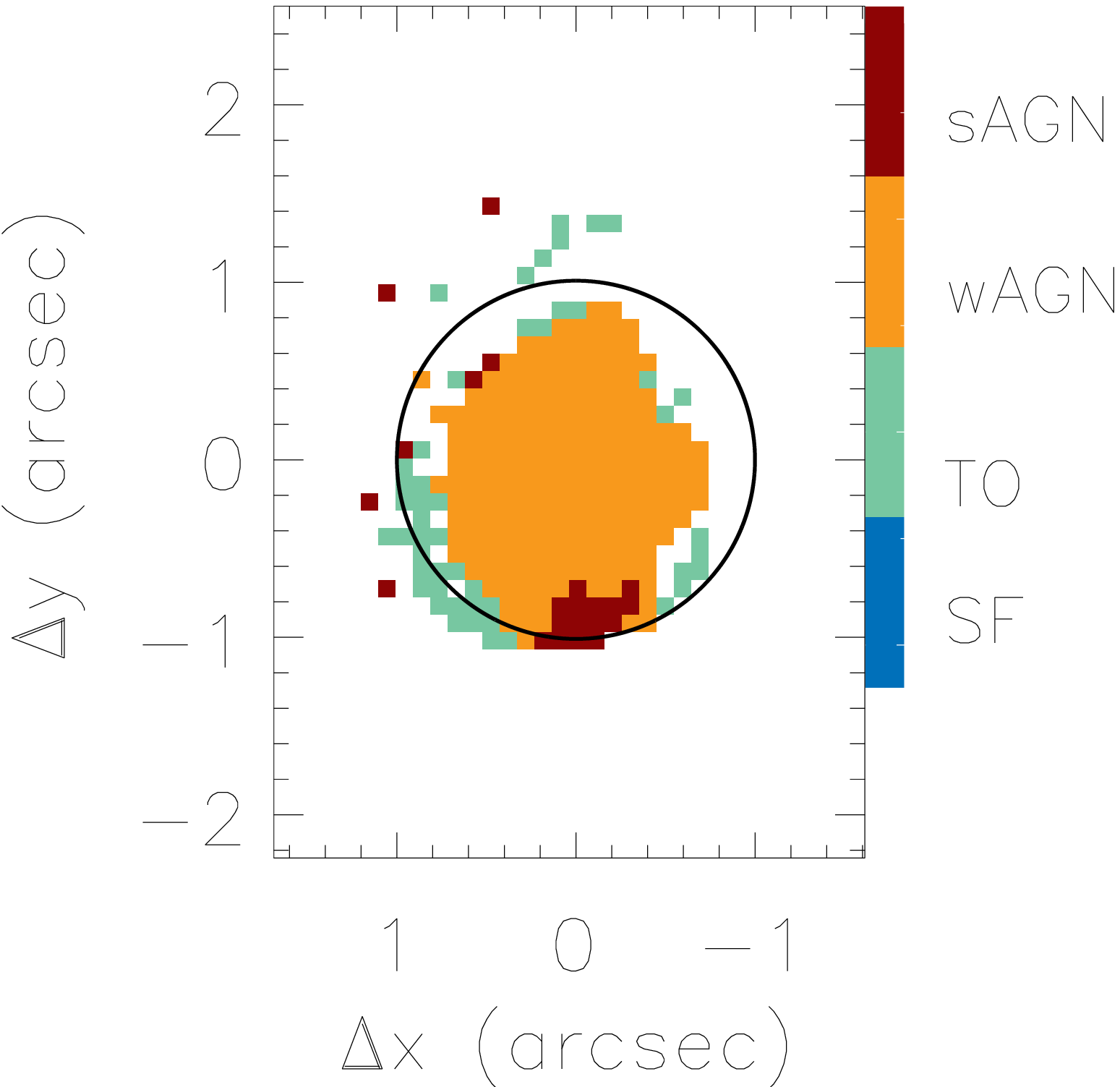} &
\includegraphics[width=0.22\textwidth]{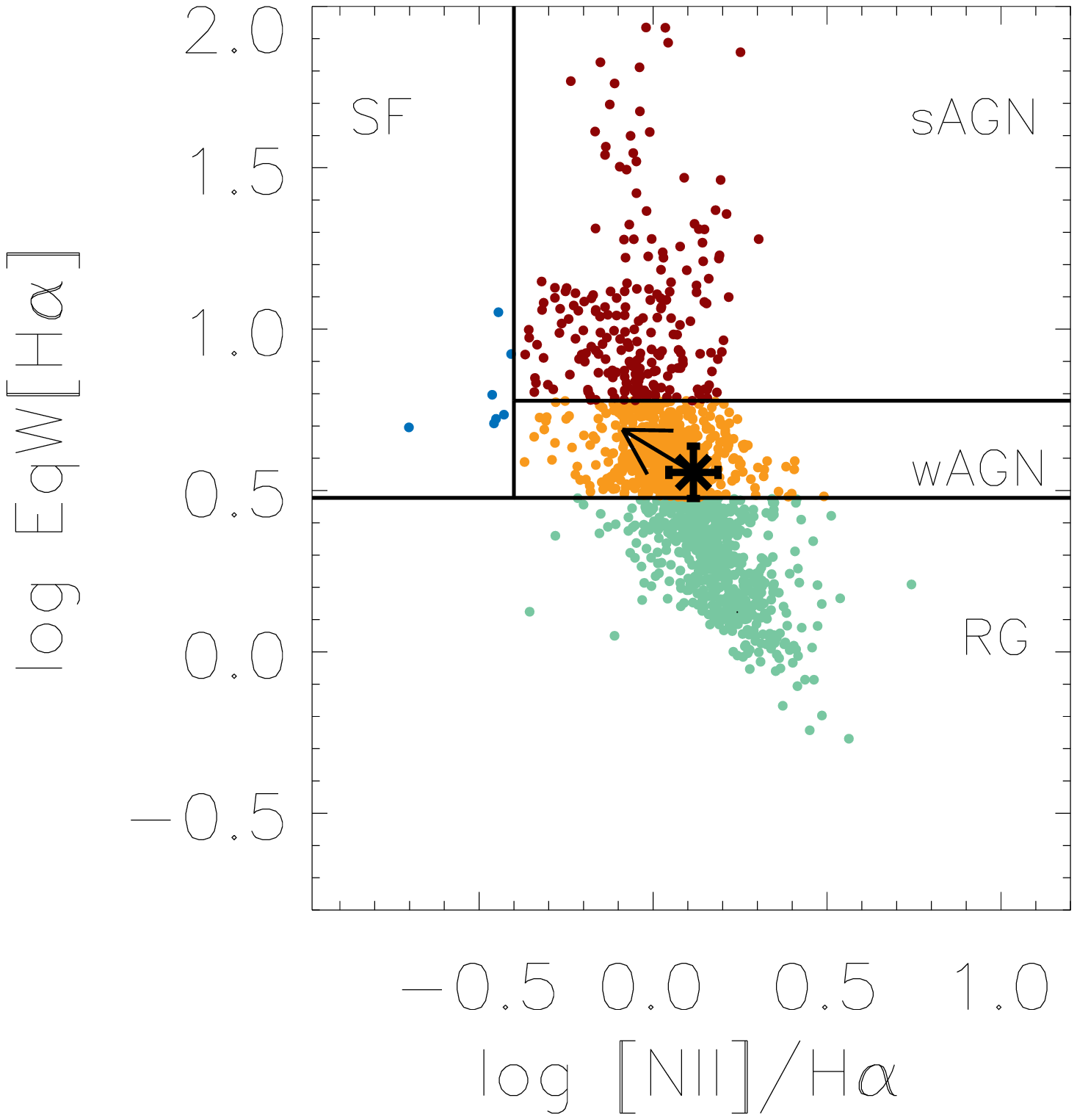} &
\includegraphics[width=0.20\textwidth]{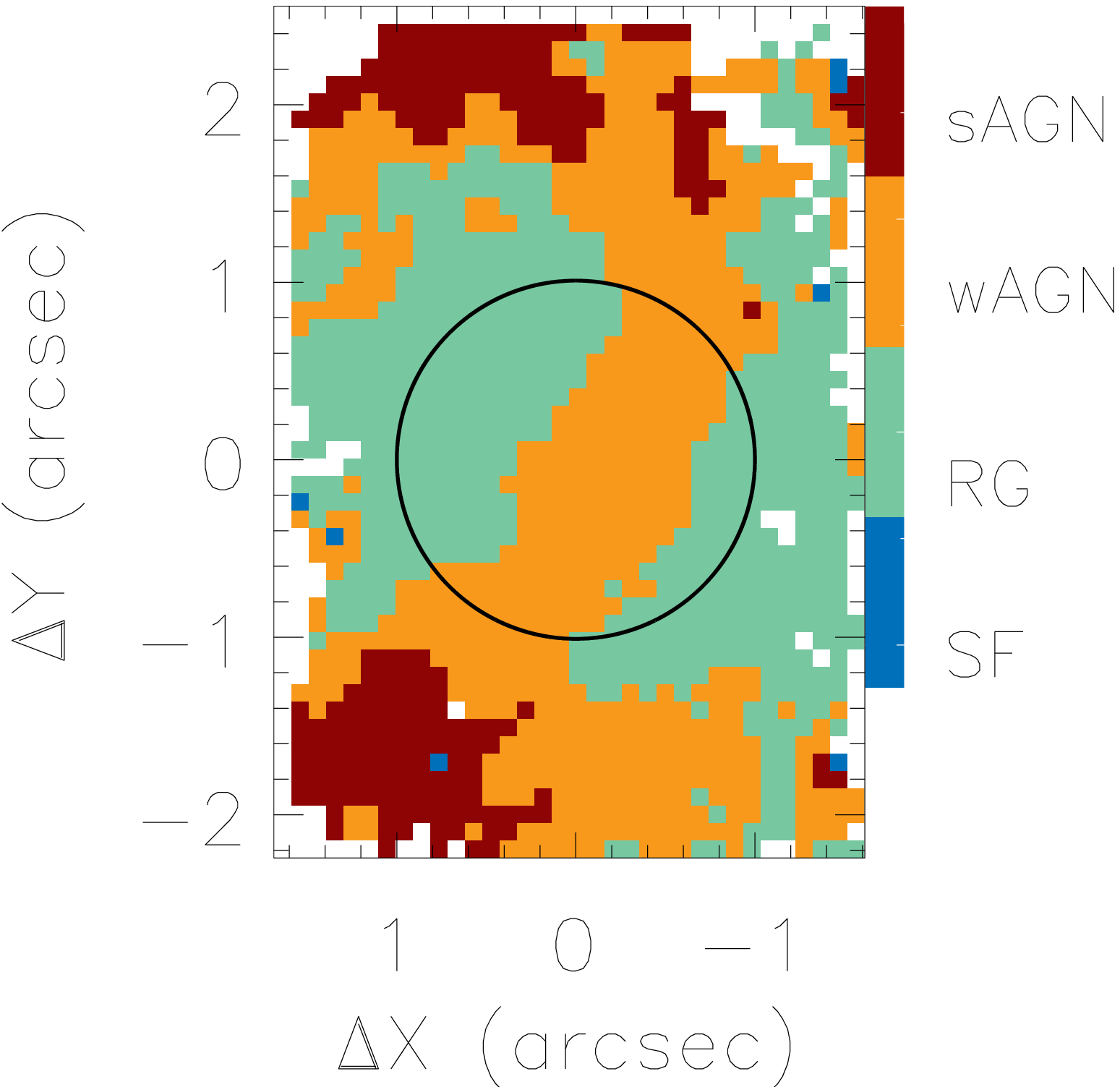} \\
  \end{tabular}
  \caption{ BPT \citep[left, ][]{bpt81} and WHAN \citep[right, ][]{cid10,cid11} diagrams from MaNGA (top) and GMOS (bottom) measurements. The color coded excitation maps show the spatial location of each excitation region shown in the BPT and WHAN diagrams. The dashed line shown in the BPT diagram is from \citet{kewley01}, the dotted line from \citet{kauffmann03} and the continuous line from \citet{cid10}. The following labels were used in the diagrams. SF: star-forming galaxies, TO: transition objects, wAGN: weak AGN (i.e. low-luminosity AGN), strong AGN (i.g. Seyferts) and RG: retired galaxies. The ``$\times$''  symbol at the BPT and WHAN diagrams correspond to the values obtained for the nucleus, as measured within an aperture of 2$^{\prime\prime}$ diameter (identified  in the excitation maps as black circles). For the GMOS diagrams, the arrows represent the measurements obtained after the subtraction of the stellar population component from the nuclear spectrum. The green rectangle overlaid to the MaNGA maps represents the GMOS FoV.}  
  \label{bpt}
\end{figure*}

\subsection{Gemini GMOS data}

Akira was observed with Gemini Multi-Object Spectrograph \citep[GMOS,][]{hook04} IFU \citep{smith02} on Dec. 25, 2017 (Project: GN-2017B-Q-26). The observations were done in the single-slit mode using the B600 grating, covering the wavelength range from 4350\,\AA\ to 7050\,\AA\ and resulting in a Field of View (FoV) of 3.5$^{\prime\prime}\times$5.0$^{\prime\prime}$. Four 1\,120 sec on-source exposures were done, resulting in a total exposure time of 1.25 hours. 


We used the IRAF
software to perform the data reduction following \citet{lena14}. 
The data cubes for each exposure were created at an angular sampling of 0\farcs1$\times$0\farcs1 and then median combined using the peak of the continuum as reference for astrometry and a sigma-clipping algorithm for bad pixel/cosmic ray removal. The average Differential Image Motion Monitor (DIMM) seeing during the observations was 0\farcs9.

\begin{figure*}
  \begin{tabular}{rl}
\includegraphics[width=0.48\textwidth]{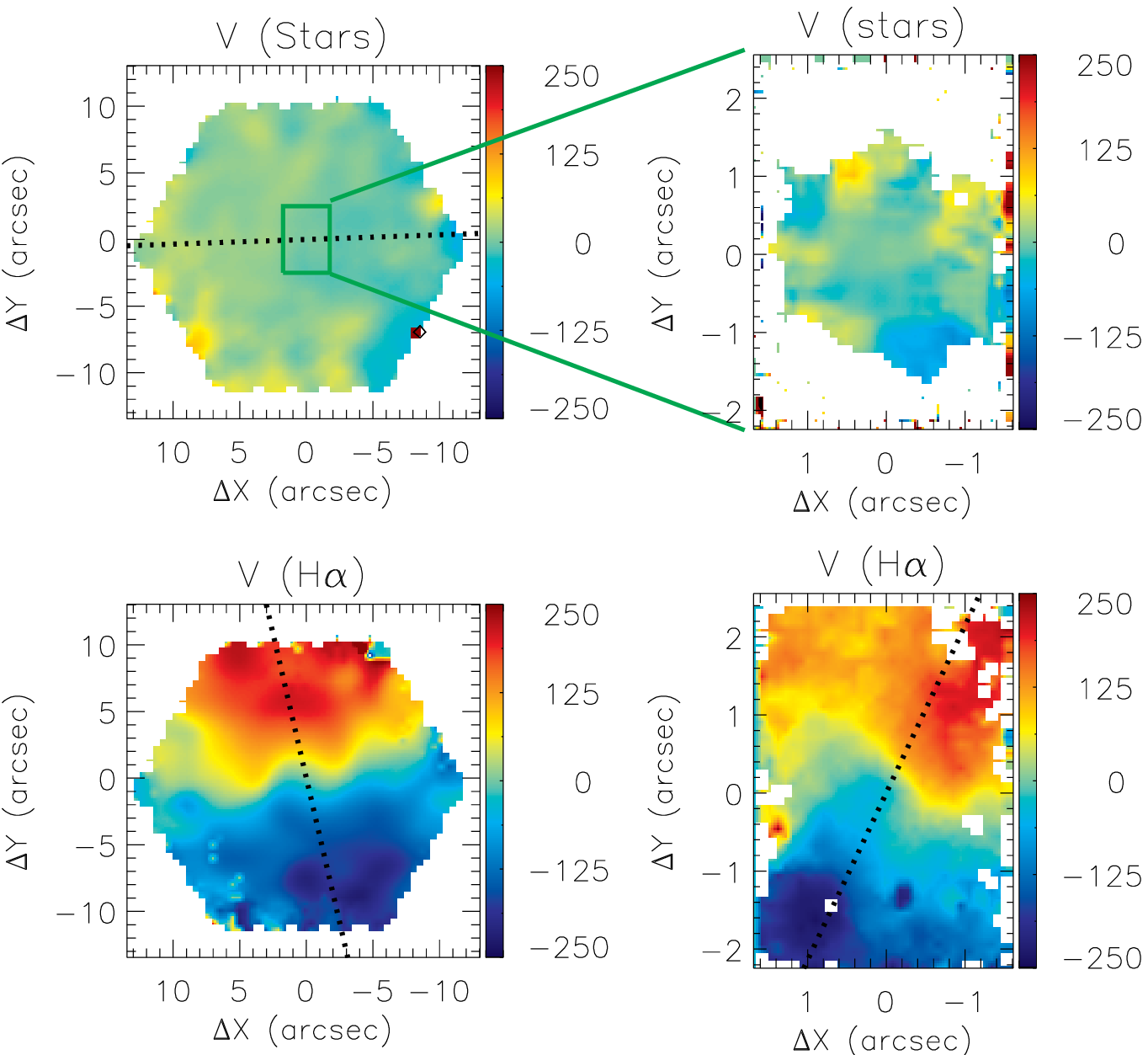} &
\includegraphics[width=0.48\textwidth]{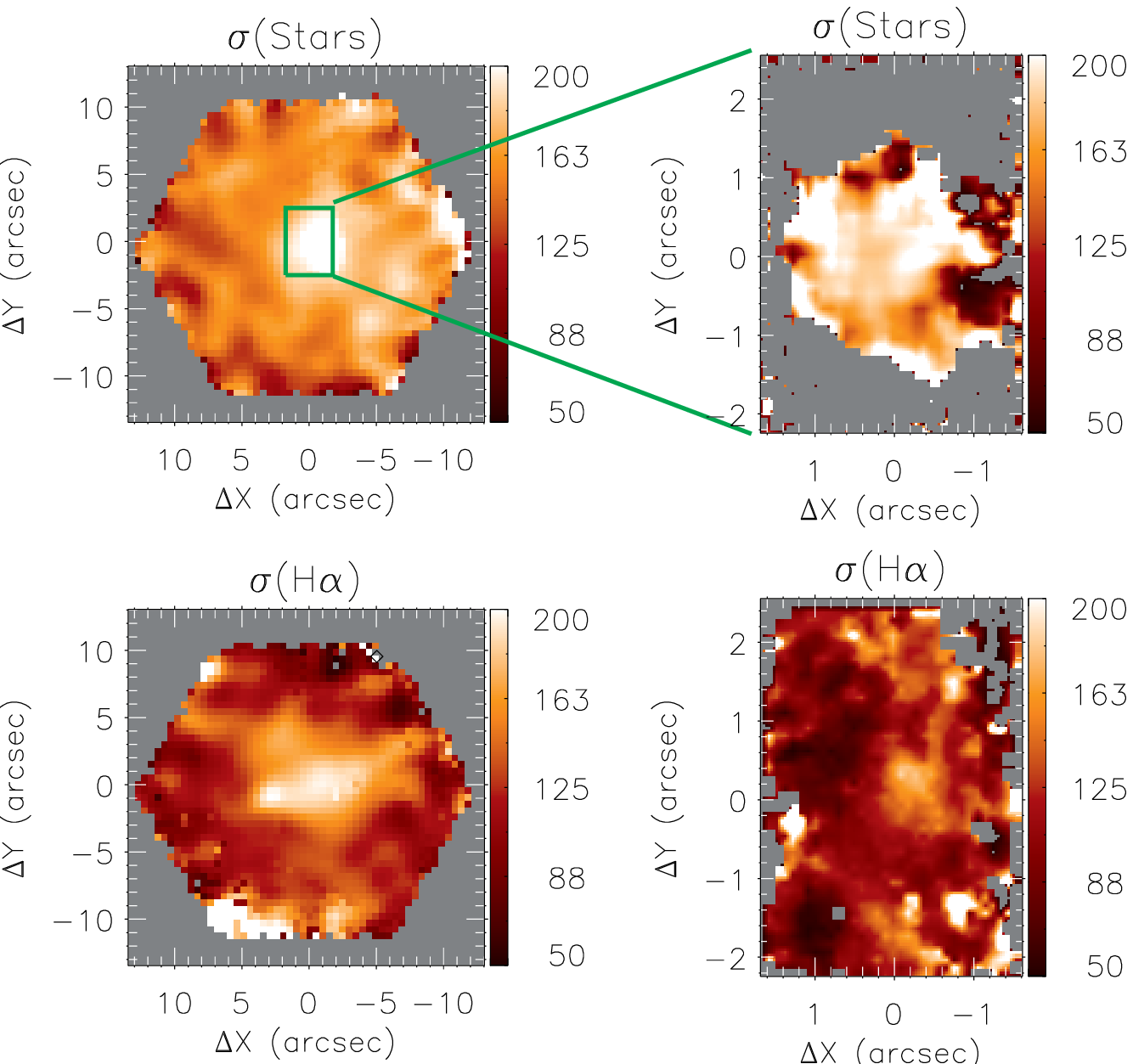} \\
  \end{tabular}
  \caption{ MaNGA and GMOS stellar and H$\alpha$ velocity fields (left panels) and $\sigma$ maps (rigth panels). The dotted lines overlaid to the velocity fields  show the orientation of the line of nodes ($\Psi_0$) as derived using the kinemetry method \citep{krajnovic06} for the whole field of view. From MaNGA data we obain $\Psi_0$=92$^\circ\pm11^\circ$ for the stars and  $\Psi_0$=13$^\circ\pm5^\circ$ for the gas. For the GMOS H$\alpha$ velocity field we obtain $\Psi_0$=155$^\circ\pm3^\circ$ and were not able to constrain the $\Psi_0$ for the stars, due to the limited spatial coverage of the measurements. The color bars show the velocity and velocity dispersion in units \kms\ after the subtraction of the systemic velocity of the galaxy and correct for the instrumental broadening, respectively. } 
  \label{kinematics}
\end{figure*}

\subsection{Emission-Line fitting}

We followed the procedure described in \citet{ilha18}, using the Gas AND Absorption Line Fitting ({\sc gandalf}) code \citep{sarzi06,oh11} to fit the emission-line profiles and the underlying stellar continuum for the MaNGA datacube, allowing us to measure simultaneously the gas and stellar kinematics. 
 The underlying stellar contribution on the galaxy spectra is fitted by GANDALF using the Penalized Pixel-Fitting ({\sc ppxf}) routine \citep{cappellari04} and requires the use of a library of template spectra. We used 30 selected Evolutionary Population Synthesis (EPS) models presented by \citet{bc03}, covering ages ranging from 5~Myr to 12~Gyr and three metallicities ($0.004\,Z_{\odot},0.02\,Z_{\odot},0.05\,Z_{\odot}$). The emission-line profiles were fitted by Gaussian curves and we kept fixed the line ratios [O\,{\sc iii}]$\lambda5007$/[O\,{\sc iii}]$\lambda4959$=$2.86$ and [N\,{\sc ii}]$\lambda6583$/[N\,{\sc ii}]$\lambda65484$=$2.94$  to their theoretical values. 

Due to the smaller spectral coverage and lower signal-to-noise ratio (S/N) of the GMOS data we were not able to use the same procedure to fit the GMOS spectra. 
Thus, we used the {\sc profit} code \citep{profit} to fit the emission-line profiles with Gaussian curves. The Gemini GMOS spectra of Akira includes the following emission lines: 
 H$\beta$, [O\,{\sc iii}]\,$\lambda\lambda$4959,5007, [O\,{\sc i}]\,$\lambda$6300, H$\alpha$, [N\,{\sc ii}]\,$\lambda\lambda$6548,83\, and [S\,{\sc ii}]\,$\lambda\lambda$6716,31, which are the most intense emission-lines from the narrow line region of AGNs.

\begin{figure*}	
\centering
\includegraphics[width=0.75\textwidth]{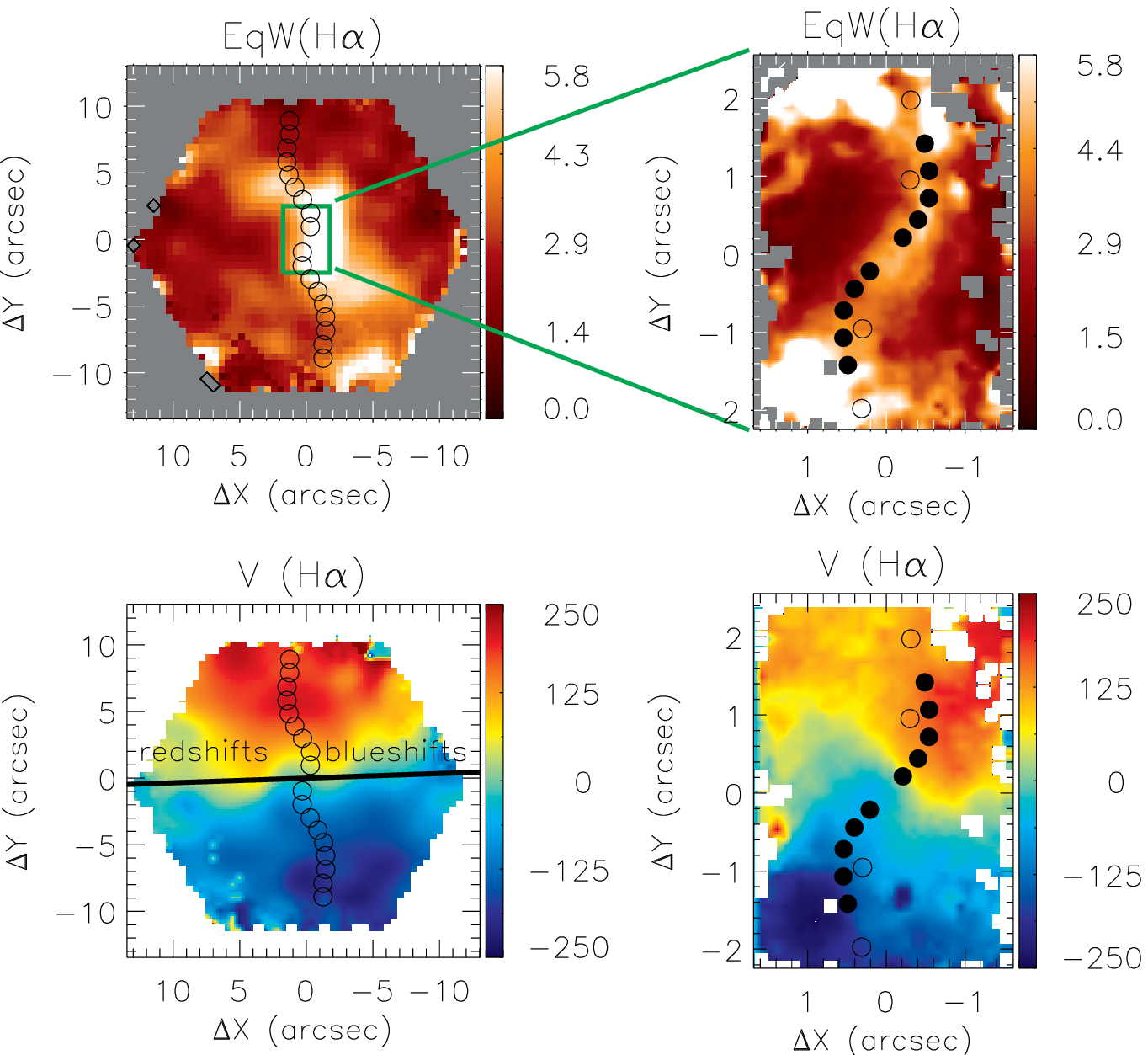} 
\includegraphics[width=0.75\textwidth]{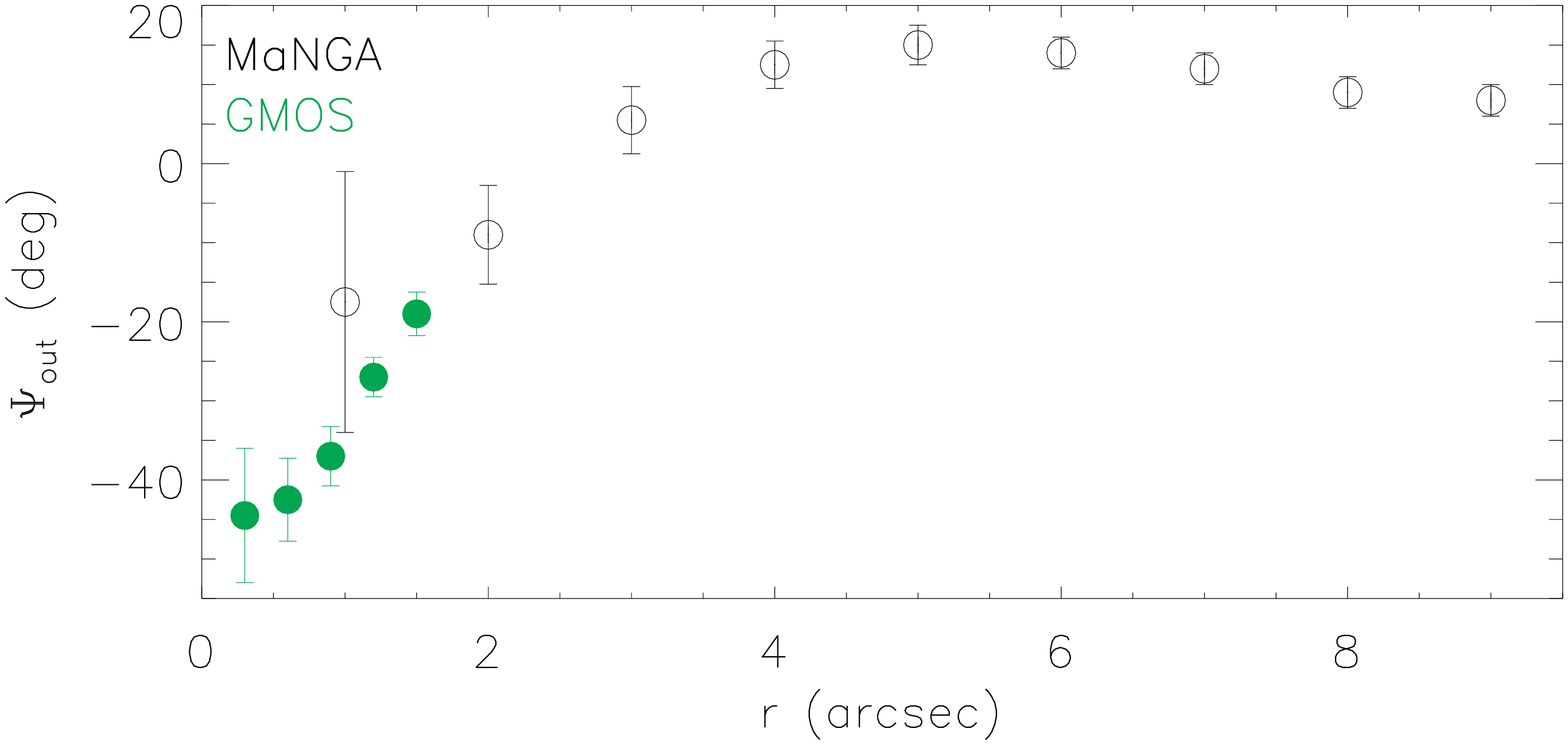}
\caption{
Large and small scale outflows in the \emph{Akira} galaxy seen with SDSS-MaNGA and Gemini. 
Top:  H$\alpha$ equivalent width (EW) maps from MaNGA (left) and Gemini (rigth) showing a stripe of larger values. \emph{Middle:} Large (left) and small (right) scale H$\alpha$ velocity fields.
The black line shows the orientation of the line of nodes measured from the stellar velocity field, which is distinct of the orientation of the outflow.  Regions where blueshifts and redshifts are seen is the stellar velocity field are labeled. The dots show the orientation of the outflow as derived using the kinemetry method. Filled circles are for Gemini GMOS data and open circles for MaNGA data. 
Bottom: Variation of the orientation of the outflow with the distance from the nucleus.
}
\label{plot_all}
\end{figure*}

\section{Results} 

Figure~\ref{flux} shows maps of the \ha\ flux  measured from the MaNGA (left) and GMOS (right) datacubes. A similar behaviour is found for other emission lines. The emission-line flux distributions at large and small scales are distinct, with the orientation of the highest intensity levels misaligned by about 50$^\circ$. Within the inner 1$^{\prime\prime}$ (0.5 kpc) the GMOS flux map shows an elongation along PA$\approx-$30$^{\circ}$, while at distances larger than 5$^{\prime\prime}$ from the nucleus, the most extended emission is seen along PA$\sim20^\circ$, as revealed by the MaNGA flux map. 

Based on the radio luminosity of Akira, \citet{cheung16} concluded that Akira harbors a LLAGN. 
Indeed, results from radio observations from VLA-FIRST data strongly suggest that most red geysers host an  AGN \citep{roy18}. We have constructed the BPT \citep{bpt81} and WHAN \citep{cid11} diagrams (Fig.~\ref{bpt}) in order to investigate the nature of the line emission in Akira.  
The WHAN diagram was originally introduced by \citet{cid10} and first used for a large sample of galaxies observed as part of the SDSS by \citet{cid11}. This diagram is able to separate sources among (i) pure star-forming galaxies with log\,[N\,{\sc ii}]/H$\alpha < -0.4$ and  Equivalent Width ($EW$) for H$\alpha$ ($EW_{H\alpha}$) $>3$\,\AA; (ii) strong AGN (sAGN; e.g. Seyfert nuclei) with log\,[N\,{\sc ii}]/H$\alpha > -0.4$ and  $EW_{H\alpha}$ $>6$\,\AA; (iii) weak AGN (wAGN; e.g. LINERS) with  log\,[N\,{\sc ii}]/H$\alpha > -0.4$ and  $3$\,\AA$<EW_{H\alpha}$ $<6$\,\AA; Retired Galaxies (RGs; i.e. fake AGN) with $<EW_{H\alpha}$ $<3$\,\AA\ and passive galaxies (lineless galaxies) with $EW<0.5$\,\AA\ for the H$\alpha$ and [N\,{\sc ii}]$\lambda6583$ emission lines.   

Both GMOS and MaNGA BPT diagrams for Akira (Fig.~\ref{bpt}) show that the emission line ratios at most locations of the galaxy fall in the region occupied by LINERs and emission-line galaxies whose ionizing photons are produced in the atmospheres of evolved
low-mass stars \citep{stasinska08,cid11,singh13,belfiore16}. Since the BPT diagram is not efficient in discriminating between these two excitation agents, the WHAN diagram has been used; it shows that most of the gas emission in Akira is photoionized by a LLAGN.
Integrating the fluxes within an aperture of 2$^{\prime\prime}$ diameter  centered at the nucleus, we find that in all diagrams the nucleus of Akira is classified as an AGN. It should be noticed that, as we are not subtracting the stellar population contribution spaxel-by-spaxel for the GMOS datacube, we may be underestimating the fluxes and Equivalent Width of the emission lines from the GMOS data, particularly important for the H lines. The arrows in the diagnostic diagrams show the location of the nucleus, if the stellar population is taken into account, as estimated by subtracting the stellar population contribution from the integrated spectrum. Thus, besides the radio emission, the observed line ratios also support the presence of a LLAGN in Akira.

The stellar and H$\alpha$ velocity fields and velocity dispersion ($\sigma$) maps of  Akira are shown in Figure~\ref{kinematics}. Considering the lower signal to noise ratio and spatial coverage of the GMOS data, we were able to measure the stellar kinematics only within the 1$^{\prime\prime}$ central. We have masked out locations where the uncertainties in velocity or $\sigma$ are larger than 25~\kms. These regions are shown in white in the stellar velocity field and in gray in the $\sigma$ map derived from the GMOS data. We show all velocity fields and all $\sigma$ maps at the same velocity scales, as labeled in the color bars. The velocity fields are shown after the subtraction of the systemic velocity of the galaxy and the $\sigma$ maps are corrected for the instrumental broadening. Although, no clear rotation pattern is observed for the stars in the inner region, the GMOS and MaNGA maps show similar range of values.

The stellar and gas kinematics of Akira measured from MaNGA data are distinct, as already noticed by \citet{cheung16}. While the amplitude of the stellar velocity field about 50~\kms, the gas velocity field reaches projected velocities of up to 250\,\kms. As already noticed by \citet{cheung16}, the kinematic major axis ($\Psi_0$) of the stellar velocity field is distinct from that of the gas velocity fields. Using the kinemetry \citep{krajnovic06} method to symmetrize the MaNGA stellar velocity, we derived $\Psi_0$=92$^\circ\pm11^\circ$, with blueshifts seen to the west and redshifts to the east. From the MaNGA H$\alpha$ velocity field, we obtain $\Psi_0$=13$^\circ\pm5^\circ$, displaced by 79$^\circ$ from the stellar value. The orientation of the kinematic major axis is indicated as dashed lines in the MaNGA stellar and H$\alpha$ velocity fields, shown in Fig.~\ref{kinematics}. In addition, the GMOS \ha\ velocity field shows that the orientation of the velocity gradient is distinct of that seen in the large scale MaNGA map. \citet{cheung16}  estimated the position angle of the major axis of Akira as $\sim53^\circ$, based on the galaxy's elliptical isophotes (from the contours in their Fig. 1c). This value is about 40$^\circ$ distinct than ours. However, it should be noticed that the isophotes of the galaxy are almost circular and thus the uncertainty in their determination of the orientation of the major axis of the galaxy may be very  high.

The $\sigma$ maps for the gas and stars show values ranging from 50 to 200 \kms, with the highest values observed at the nucleus of the galaxy. In the inner region, the $\sigma$ values measured from MaNGA and GMOS data are consistent each other. For the gas, the GMOS map shows a `S-shaped' structure of higher values that seems to follow the structure of highest H$\alpha$ emission seen in Fig.~\ref{flux} and the higher ionization structure seen in the WHAN diagram of Fig.~\ref{bpt}.

\section{Discussion}

\subsection{Gas kinematics}

As already mentioned in previous section, we derive that the orientation of the line of nodes for the H$\alpha$ velocity field is $\Psi_0$=13$^\circ\pm5^\circ$, as obtained using the  kinemetry \citep{krajnovic06} method applied to the MaNGA data. We note that the velocity gradient in the inner region of Akira is distinct to that seen on large scales (see Fig.~\ref{plot_all}). Using the kinemetry method for the GMOS H$\alpha$ velocity field  we obtain that the main gradient is observed along  $\Psi_0$=155$^\circ\pm3^\circ$. 

\begin{figure}	
\centering
\includegraphics[width=0.48\textwidth]{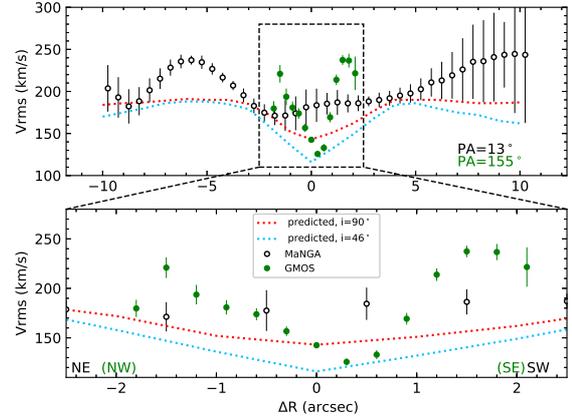} 
\caption{$V_{\rm rms}$ as obtained from MaNGA (open circles) and GMOS (closed circles) measurements, extracted along the orientation of the main velocity gradient seen in H$\alpha$ velocity field. The $V_{\rm rms}$ were calculated within a pseudo-slit with width of 0\farcs5 oriented along PA$=10^ \circ$ for MaNGA data and 155$^\circ$ for GMOS data. The red and blue dotted curves show the predicted $V_{\rm rms}$ values by \citet{cheung16} for disk inclinations of 90$^\circ$ and 46$^\circ$, respectively. 
}
\label{vrms}
\end{figure}

In order to verify if the observed velocity fields are consistent with motions dominated by the gravitational potential of the galaxy we have derived the second velocity moment $V_{\rm rms}=\sqrt{V^2+\sigma^2}$, where $V$ is the line centroid velocity and $\sigma$ its velocity dispersion.  In Figure~\ref{vrms} we present the plot of $V_{\rm rms}$ as a function of the distance from the nucleus. The $V_{\rm rms}$ was computed using the H$\alpha$ velocity and $\sigma$ measurements within a pseudo-slit with width of 0\farcs5 and oriented along the main velocity gradients observed in the MaNGA (13$^\circ$) and GMOS (155$^\circ$) H$\alpha$ velocity fields.  The dotted curves are from \citet{cheung16} and represent the  predicted $V_{\rm rms}$ values by Jeans Anisotropic Modelling \citep[JAM; ][]{jam} for two disk inclinations ($i$). The red curve corresponds to the predicted values for $i=90^\circ$ and the blue line shows the predicted  $V_{\rm rms}$ for $i=46^\circ$, t the minimum allowed axial ratio derived by the authors using  GALFIT fits of the H$\alpha$ flux distribution. More details about the dynamical modeling can be found in \citet{cheung16}. As can be observed in Fig.~\ref{vrms} and already discussed by \citet{cheung16}, the observed $V_{\rm rms}$ for MaNGA data exceed by up to $\sim$100\,km,s$^{-1}$ the predicted values for disk rotation.  In the inner region, a similar behaviour is observed for the GMOS data. For distances above 0\farcs5 from the nucleus, the observed values are more than 50\,km\,s$^ {-1}$ higher than the dynamical model predictions.  

Thus, considering that $V_{\rm rms}$ values derived from higher resolution GMOS data follows the same behaviour of the large scale MaNGA data and exceed the predicted $V_{\rm rms}$ values by $\gtrsim50$\,km\,s$^ {-1}$, we conclude that not only the large scale gas kinematics, but also the kinematics in the inner few hundred parsecs cannot be described by disk-like rotation. Thus, as already concluded by \citet{cheung16}, the most probably explanation for the observed gas kinematics of Akira is that it is due to centrally driven outflows. 

\subsection{The orientation of the outflow}

In Figure \ref{plot_all} we present the H$\alpha$ EW maps obtained from MaNGA (left panel) and GMOS (right panel) data. 
As for the flux maps shown in Fig.~\ref{flux}, the EW maps clearly show that the large and small scale structures are misaligned. The middle panels of Fig. \ref{plot_all} show the large (left) and small (right) scale H$\alpha$ velocity fields.
\citet{cheung16} concluded that the large scale gas kinematics observed for Akira cannot be explained by motions of the gas due to the gravitational potential of the galaxy, being most probably due to outflows from a central AGN.

In order to better investigate the gas kinematics and the origin of the suggested outflow in Akira, we have used the kinemetry method to measure also the variation of  orientation of the outflow ($\Psi_{\rm out}$)  with the distance from the nucleus using the H$\alpha$ velocity field, as it is the strongest emission line at most locations. For the GMOS data, we applied the kinemetry method for concentric rings of $r_w$=0\farcs6 width by varying its distance to the nucleus in bins of $dr$=0\farcs3. For the MaNGA data, we adopted  $r_w$=2\farcs0 and $dr$=1\farcs0. These values were chosen based on the angular resolution of the data, so that in the inner disk the bin samples better the resolution of the data. In the bottom panel of Fig.~\ref{plot_all}, we present a plot of the orientation of the outflow $\Psi_{\rm out}$ vs. the distance to the nucleus. 
We note that the orientation of the outflow changes from $\Psi_{\rm out}\approx-50^\circ$ at the nucleus to $\Psi_{\rm out}\approx15^\circ$ at 5$^{\prime\prime}$ from it. At larger distances, the value of $\Psi_{\rm out}$ shows a slight decrease.

\subsection{The origin of the outflow}

We have reported the detection of a varying orientation in the wind launched from the nucleus of Akira, as a function of distance from it.
The intensity-line ratio diagnostic diagrams (Figure~\ref{bpt}) confirm that Akira hosts an AGN, therefore favoring the idea that the wind is launched by the SMBH rather than being supernovae-driven. In this section, we discuss the possible origins of the outflow and conjecture on the nature of the observed precession. 


 Akira shows point-like, core radio emission detected with VLA \citep{roy18} and seems to be radio-loud, much like other LLAGNs \citep{Ho02, younes12}. Since it does not display evidence for the presence of any kpc-scale radio emission, Akira does not seem to produce extended relativistic jets like radio galaxies. For this reason, we think  the outflow that we are tracing with MANGA and Gemini IFU optical observations is associated with a subrelativistic wind which spreads its power over a larger surface area than a collimated jet. In fact, such winds are expected to be a natural outcome of SMBHs accreting in the RIAF mode, as suggested by numerous theoretical works \citep[e.g.][]{begelman12,sadowski13,yuan15,bu16}.

Having established the likely wind-nature of the kpc-scale outflows observed in Akira, we now turn to the origin of the observed precession. First of all, most SMBHs should have some degree of angular momentum \citep{volonteri05,King08}, even though it is not know how high the spin parameter is in the LLAGN population \citep{Reynolds13}. Secondly, there is no reason why the angular momentum vectors of the gas fed to the SMBH at large distances should know about the SMBH spin vector. Therefore, we should expect a natural misalignment between the accretion disk and BH spin leading to a torque exerted by the BH which introduces a global precession in the disk--the Lense-Thirring precession \citep{Bardeen1975,Fragile2007}. The precession should be manifested in any outflows originating from the disk due to angular momentum conservation \citep[][Liska, Fragile, private communication]{Liska2018}. We conclude that Lense-Thirring precession is a possible origin for the wind precession observed in Akira at kpc-scales. The change in the orientation in the ionized gas pattern (as seen in the EW maps of Fig.~\ref{plot_all}) also supports the precession of accretion disk.  

We can use a simple argument to estimate the launching radius of the wind, assuming Lense-Thirring precession. 
From angular momentum conservation, we have
\begin{equation}
m \Omega r^2 = {\rm constant}
\end{equation}
throughout the wind, where $m$ is the mass of a gas element, $\Omega$ is the wind precession angular velocity and $r$ is the distance to the SMBH. We apply the above equation to two separate regions of the flow--the footpoint in the accretion disk where the wind is launched and the location probed by the Gemini IFU observations at kpc-scales, from which it follows that
\begin{equation}	\label{disklaunch}
\Omega_d r_d^2 = \Omega_w r_w^2,
\end{equation}
where the $d$ and $w$ subscripts refer to either the accretion disk or the wind. The above equation relates the measured wind properties at kpc-scales to the conditions closer to the black hole. 

We can use equation \ref{disklaunch} to estimate the distance from the SMBH at which the wind is launched, $r_d$. In order to do so, we need estimates of $\Omega_d$, $\Omega_w$ and $r_w$. From the lower panel of Fig. \ref{plot_all}, we see that the wind completed a precession of about $\Delta \phi_w = 50^\circ$. We can obtain an order-of-magnitude estimate of the time it takes to precess by this angle as the time it takes the wind to reach a distance of 5 arcsec from the SMBH (about 2.5 kpc). Using the current wind velocity as estimated by \cite{cheung16} of 310 km/s, the outflow will reach the above distance in $\Delta t_w < 10^7$ yrs., assuming that the gas slowed down compared to its velocity at the launching point. We estimate, then, 
\begin{equation}
\Omega_w = \frac{\Delta \phi_w}{\Delta t_w} \sim 10^{-7} \ {\rm rad \ yr}^{-1} 
\end{equation}
at $r_w = 2.5$ kpc. The calculation of the  disk's $\Omega_w$ is model-dependent; from numerical, general relativistic magnetohydrodynamic simulations of tilted accretion disks, the disk precession frequency due to the Lense-Thirring effect is 
\begin{equation}
\Omega_d \approx 2\pi \left( \frac{M}{10^8 M_\odot} \right)^{-1} \ {\rm rad \ yr}^{-1}
\end{equation}
where $M$ is the SMBH mass \citep{Fragile2007}. Plugging in the above estimates of $\Omega_w$, $\Omega_d$ and $r_w$, equation \ref{disklaunch} gives us a launching radius of 
\begin{equation}
r_d \sim (0.1-1) \ {\rm pc} \sim  \left( 10^4-10^5 \right) R_S
\end{equation}
where $R_S \equiv 2GM/c^2$ is the Schwarzschild radius, for a BH mass of $10^8 M_\odot$ as appropriate for Akira. This launching radius is 2-3 orders of magnitude larger than expected from current theories of thermally-driven or magnetically-driven winds from RIAFs ($\sim 10-100 R_S$; \citealt{sadowski13,yuan15}). This disagreement can be alleviated if the BH mass is much larger than $10^8 M_\odot$ or if the disk precession frequency is considerably faster than current predictions of tilted, thick accretion disk models.

Another possible origin of precession of a wind could the Bardeen-Peterson precession \citep{Bardeen1975, Caproni2006} from a previous thin disk. This would be the case if Akira's nucleus was brighter in the past and its SMBH was correspondingly accreting at higher rates, and only more recently became underfed and turned into a LLAGN -- similarly to Hanny's Voorwerp and IC 2497 \citep{Lintott2009,Keel2012}. This scenario is supported by the classical AGN feedback picture \citep{heckman14}, where the ``quasar mode" of accretion triggered by violent mergers halt the star formation, and when the quasar phase turns off, the ``maintenance mode'' comes into effect from the underfed AGN and they become LLAGN.  

Irrespective of the physical nature of the precession, when occurring on a wind it can increase its working surface, such that the wind can spread its power over a larger area and stir up the ambient gas much more effectively than a narrow, collimated jet -- i.e. precessing winds make AGN feedback more efficient for quenching star formation  \citep[e.g.,][]{Falceta-Goncalves2010}. 


\section{Conclusions}

We used IFS obtained with the Gemini GMOS-IFU and SDSS-IV--MaNGA to investigate the gas kinematics of the galaxy Akira -- the prototypical ``Red Geyser''. We found that the outflow can be resolved with GMOS-IFU down to the nucleus and changes orientation by about $50^\circ$ from the nucleus of the galaxy to kiloparsec scales on a timescale of $10^7$ years, being spatially correlated with the complex structure of high EW emission-line values. These observations are consistent with an origin in a precessing accretion disk. Precession in wide, sub-relativistic outflows can help to spread their power over a larger area and increase the effectiveness of AGN feedback in galaxies similar to Akira. This could explain how a relatively low-power wind is nonetheless able to efficiently quench star formation and maintain the quiescence in typical galaxies. 


\section*{Acknowledgements}
We thank an anonymous referee for valuable suggestions which helped to improve the paper and 
 acknowledge useful discussions with Chris Fragile, Matthew Liska and Anderson Caproni. We thank an anonymous referee for valuable suggestions which helped to improve the paper.
This study was financed in part by the Coordena\c c\~ao de
Aperfei\c coamento de Pessoal de N\'ivel Superior - Brasil (CAPES) -
Finance Code 001, Conselho Nacional de Desenvolvimento Cient\'ifico e Tecnol\'ogico (CNPq) and Funda\c c\~ao de Amparo \`a pesquisa do Estado do RS (FAPERGS). RN acknowledges funding by the Funda\c{c}\~ao de Amparo \`a Pesquisa do Estado de S\~ao Paulo (FAPESP) through grant 2017/01461-2.

Based on observations obtained at the Gemini Observatory, which is operated by the Association of Universities for Research in Astronomy, Inc., under a cooperative agreement with the NSF on behalf of the Gemini partnership: the National Science Foundation (United
States), the Science and Technology Facilities Council (United Kingdom), the
National Research Council (Canada), CONICYT (Chile), the Australian Research Council
(Australia), Minist\'erio da Ci\^encia e Tecnologia (Brazil) 
and Ministerio de Ciencia, Tecnolog\'ia e Innovaci\'on Productiva  (Argentina).

SDSS is managed by the Astrophysical Research Consortium for the Participating Institutions of the SDSS Collaboration including the Brazilian Participation Group, the Carnegie Institution for Science, Carnegie Mellon University, the Chilean Participation Group, the French Participation Group, Harvard-Smithsonian Center for Astrophysics, Instituto de Astrofisica de Canarias, The Johns Hopkins University, Kavli Institute for the Physics and Mathematics of the Universe (IPMU) / University of Tokyo, the Korean Participation Group, Lawrence Berkeley National Laboratory, Leibniz Institut f\"ur Astrophysik Potsdam (AIP), Max-Planck-Institut f\"ur Astronomie (MPIA Heidelberg), Max-Planck-Institut f\"ur Astrophysik (MPA Garching), Max-Planck-Institut f\"ur Extraterrestrische Physik (MPE), National Astronomical Observatories of China, New Mexico State University, New York University, University of Notre Dame, Observat\' orio Nacional / MCTI, The Ohio State University, Pennsylvania State University, Shanghai Astronomical Observatory, United Kingdom Participation Group, Universidad Nacional Aut\'onoma de M\'exico, University of Arizona, University of Colorado Boulder, University of Oxford, University of Portsmouth, University of Utah, University of Virginia, University of Washington, University of Wisconsin, Vanderbilt University, and Yale University.




\begin{thebibliography}{99}
\bibitem[\protect\citeauthoryear{Albareti et al.}{2017}]{albareti17} Albareti, F. D. et al. 2017, ApJS, 233, 285.
\bibitem[\protect\citeauthoryear{Antonucci}{1993}]{antonucci93} Antonucci, R., 1993, ARAA, 31, 473.
\bibitem[\protect\citeauthoryear{Baldwin, Phillips \& Terlevich}{1981}]{bpt81}
Baldwin, J. A., Phillips, M. M., Terlevich, R., 1981, PASP, 93, 5.
\bibitem[\protect\citeauthoryear{Barbosa et al.}{2014}]{barbosa14} Barbosa, F. K. B. et al., 2014, MNRAS, 445, 2353.
\bibitem[\protect\citeauthoryear{Bardeen \& Petterson}{1975}]{Bardeen1975} Bardeen, J. M.; Petterson, J. A. 1975, ApJ, 195, 65.
\bibitem[\protect\citeauthoryear{Belfiore et al.}{2016}]{belfiore16} Belfiore, F., et al. 2016, MNRAS, 461, 3111.
\bibitem[\protect\citeauthoryear{Benson et al.}{2003}]{benson03} Benson, A. J., Bower, R. G., Frenk, C. S., Lacey, C. G., Baugh, C. M., Cole, S., 2003, ApJ, 599, 38.
\bibitem[\protect\citeauthoryear{Blanton et al.}{2017}]{blanton17} Blanton, M. R. et al. 2017, AJ, 154, 28.
\bibitem[\protect\citeauthoryear{Bower et al.}{2006}]{bower06} Bower, R. G. et al. 2006, MNRAS, 370, 645.
\bibitem[\protect\citeauthoryear{Bruzual \& Charlot}{2003}]{bc03}  Bruzual G., Charlot S., 2003,   MNRAS,  344, 1000
\bibitem[\protect\citeauthoryear{Bundy et al.}{2015}]{bundy15} Bundy, K. et al. 2015, ApJ, 798, 7.
\bibitem[\protect\citeauthoryear{Cappellari \& Emsellem}{2004}]{cappellari04} \bibitem[\protect\citeauthoryear{Cappellari}{2008}]{jam} Cappellari, M., 2008, MNRAS, 390, 71.
\bibitem[\protect\citeauthoryear{Caproni et al.}{2006}]{Caproni2006} Caproni, A.; Abraham, Z.; Mosquera Cuesta, H. J., 2006, ApJ, 638, 120.
\bibitem[\protect\citeauthoryear{Caproni et al.}{2013}]{Caproni2013} Caproni, A.; Abraham, Z.; Monteiro, H. J., 2013, MNRAS, 428, 280.
\bibitem[\protect\citeauthoryear{Cattaneo et al.}{2009}]{cattaneo09} Cattaneo, A. et al., 2009, Nature, 460, 213.
\bibitem[\protect\citeauthoryear{Cid Fernandes et al.}{2010}]{cid10}
Cid Fernandes, R., Stasi\'nska, G., Schlickmann, M. S., Mateus, A., Vale Asari, N., Schoenell, W., Sodr\'e, L., 2010, MNRAS, 403, 103
\bibitem[\protect\citeauthoryear{Cid Fernandes et al.}{2011}]{cid11} Cid Fernandes, R., Stasi\'nska, G., Mateus, A., Vale Asari, N., 2011, MNRAS, 413, 1687.
\bibitem[\protect\citeauthoryear{Cheung et al.}{2016}]{cheung16} Cheung, E., et al., 2016, Nature, 533, 504.
\bibitem[\protect\citeauthoryear{Cresci et al.}{2015}]{cresci15}  Cresci, G. et al. 2015, A\&A, 582, A63.
\bibitem[\protect\citeauthoryear{Di Matteo, Springel, \& Hernquist}{2005}]{dimateo05} Di Matteo, T., Springel, V. \& Hernquist, L. 2005, Nature, 433, 604.
\bibitem[\protect\citeauthoryear{Falceta-Gon\c calves et al.}{2010}]{Falceta-Goncalves2010} Falceta-Gon\c calves, D.; Caproni, A.; Abraham, Z.; Teixeira, D. M.; de Gouveia Dal Pino, E. M., 2010, ApJ, 713, 74.
\bibitem[\protect\citeauthoryear{Drory et al.}{2015}]{drory15} Drory, N., et al. 2015, AJ, 149, 77.  
\bibitem[\protect\citeauthoryear{Fischer et al.}{2013}]{fischer13} Fischer, T. C.  et al., 2013, ApJS, 209, 1.
\bibitem[\protect\citeauthoryear{Fragile et al.}{2007}]{Fragile2007} Fragile, P. C.; Blaes, O. M.; Anninos, P.; Salmonson, J. D., 2007, ApJ, 664, 417.	
\bibitem[\protect\citeauthoryear{Gunn et al.}{2006}]{gunn06} Gunn, J. E., et al. 2006, AJ, 131, 2332           
\bibitem[\protect\citeauthoryear{Harrison et al.}{2018}]{harrison18} Harrison, C. M.; Costa, T.; Tadhunter, C. N.; Flutsch, A.; Kakkad, D.; Perna, M.; Vietri, G., 2018, NatAs, 2, 198. 
\bibitem[\protect\citeauthoryear{Heckman \& Best}{2014}]{heckman14} Heckman, T. M., Best, P. N., 2014, ARA\&A, 52, 589.	
\bibitem[\protect\citeauthoryear{Hook et al.}{2004}]{hook04} Hook, I., Jorgensen, I., Allington-Smith, J. R., Davies, R. L., Metcalfe, N., Murowinski, R. G., Crampton, D., 2004, PASP, 116, 425.
\bibitem[\protect\citeauthoryear{Hopkins et al.}{2005}]{hopkins05}  Hopkins, P. et al. 2005, ApJ, 630, 705.
\bibitem[\protect\citeauthoryear{Ilha et al.}{2018}]{ilha18}  Ilha, G. et al. 2018, MNRAS, submitted.
\bibitem[\protect\citeauthoryear{Karouzos et al.}{2016a}]{karouzos16a} Karouzos, M., Woo, J-H., Bae, H-J., 2016a, ApJ, 819, 148. 
\bibitem[\protect\citeauthoryear{Karouzos et al.}{2016b}]{karouzos16b} Karouzos, M., Woo, J-H., Bae, H-J., 2016b ApJ, 833, 171;
\bibitem[\protect\citeauthoryear{Kauffmann et al.}{2003}]{kauffmann03}  
Kauffmann G., Heckman T.~M., Tremonti C., et al., 2003, MNRAS,  346, 1055 
\bibitem[\protect\citeauthoryear{Keel et al.}{2012}]{Keel2012} Keel, W. C. et al. , 2012, ApJ, 144, 66.
\bibitem[\protect\citeauthoryear{Kewley et al.}{2001}]{kewley01}
Kewley  L.  J.,  Dopita  M.  A.,  Sutherland  R.  S.,  Heisler  C.  A., Trevena J., 2001, ApJ, 556, 121
\bibitem[\protect\citeauthoryear{Krajnovi\'c et al.}{2006}]{krajnovic06}  Krajnovi\'c, D., Cappellari, M., de Zeeuw, P. T., Copin, Y., 2006, MNRAS, 366, 787.
\bibitem[\protect\citeauthoryear{Law et al.}{2015}]{law15} Law, D.R., et al. 2015, AJ, 150, 19   
\bibitem[\protect\citeauthoryear{Law et al.}{2016}]{law16} Law D. R., et al., 2016, AJ, 152, 83
\bibitem[\protect\citeauthoryear{Lena}{2014}]{lena14} Lena, D., 2014, arXiv:1409.8264
\bibitem[\protect\citeauthoryear{Lena et al.}{2015}]{lena15} Lena, D. et al., 2015, ApJ, 806, 84.
\bibitem[\protect\citeauthoryear{Lintott et al.}{2009}]{Lintott2009} Lintott, C. J. et al., 2009, MNRAS, 399, 129.
\bibitem[\protect\citeauthoryear{Liska et al.}{2018}]{Liska2018} Liska, M.; Hesp, C.; Tchekhovskoy, A.; Ingram, A.; van der Klis, M.; Markoff, S., 2018, MNRAS, 474, 81.
\bibitem[\protect\citeauthoryear{Oh et al.}{2011}]{oh11} Oh, K., Sarzi, M., Schawinski, K., Yi, S. K., 2011, ApJS, 195, 130.
\bibitem[\protect\citeauthoryear{Penny et al.}{2018}]{penny18} Penny, S., 2018, MNRAS, 476, 979.
\bibitem[\protect\citeauthoryear{Pope et al.}{2009}]{pope09} Pope, E. C. D., 2009, MNRAS, 395, 2317.

\bibitem[\protect\citeauthoryear{Riffel, Storchi-Bergmann \& Rifffel}{2014}]{riffel14}  Riffel, R. A., Storchi-Bergmann, T., Riffel, R., 2014, ApJ, 780, 24.
\bibitem[\protect\citeauthoryear{Riffel}{2010}]{profit} Riffel, R. A., 2010, Ap\&SS, 327, 239.
\bibitem[\protect\citeauthoryear{Riffel et al.}{2018}]{rogemar18} Riffel, R. A., Hekatelyne, C., Freitas, I. C., 2018, PASA, 35, 40.
\bibitem[\protect\citeauthoryear{Roy et al.}{2018}]{roy18} Roy, N., 2018, ApJ, accepted.

\bibitem[\protect\citeauthoryear{Sarzi at al.}{2006}]{sarzi06} Sarzi, M., et al, 2006, MNRAS, 366, 1151.
\bibitem[\protect\citeauthoryear{Schmitt et al.}{2003}]{schmitt03} Schmitt, H. R. et al., 2003, ApJS, 148, 327.
\bibitem[\protect\citeauthoryear{Schnorr-M\"uller et al.}{2014}]{sm14} Schnorr-M\"uller, A., et al., 2014, MNRAS, 437, 1708.
\bibitem[\protect\citeauthoryear{Singh et al.}{2013}]{singh13} Singh, R. et al., 2013, A\&A, 558, 43.
\bibitem[\protect\citeauthoryear{Smee et al.}{2013}]{smee13} Smee, S. A. et  al. 2013, AJ, 146, 32
\bibitem[\protect\citeauthoryear{Smith et al.}{2002}]{smith02} Smith, J. et al., 2002, PASP, 114, 892.
\bibitem[\protect\citeauthoryear{Springel, Di Matteo,  \& Hernquist}{2005}]{springel05} Springel, V., Di Matteo, T., \& Hernquist, L. 2005, ApJ, 620, L79.
\bibitem[\protect\citeauthoryear{Stasi\'nska et al.}{2008}]{stasinska08} Stasi\'nska, G. et al., 2008, MNRAS, 391, 29
\bibitem[\protect\citeauthoryear{Urry \& Padovani}{1995}]{urry95} Urry, C. M., Padovani, P., 1995, PASP, 107, 803.
\bibitem[\protect\citeauthoryear{Venturi et al.}{2018}]{venturi18}  Venturi, G. et al. 2018, A\&A, 619, A74.
\bibitem[\protect\citeauthoryear{Yan et al.}{2016a}]{yan16a} Yan R., et al., 2016, AJ, 151, 8.
\bibitem[\protect\citeauthoryear{Wake et al.}{2017}]{wake17} Wake, D.A., et al. 2017, AJ, 154, 86 
\bibitem[\protect\citeauthoryear{Yean et al.}{2016b}]{yan16b} Yan R., et al., 2016, AJ, 152, 197

\bibitem[\protect\citeauthoryear{Begelman, M. C.}{2012}]{begelman12} Begelman, M. C., 2012, MNRAS , 420 , 2912.

\bibitem[\protect\citeauthoryear{Bu, D.-F.; Yuan, F.; Gan, Z.-M. \& Yang, X.-H.}{2016}]{bu16} Bu, D.-F.; Yuan, F.; Gan, Z.-M. \& Yang, X.-H., 2016, APJ , 823 , 90.

\bibitem[\protect\citeauthoryear{Ho, L. C.}{2002}]{Ho02} Ho, L. C., 2002, APJ, 564, 120.

\bibitem[\protect\citeauthoryear{King, A. R.; Pringle, J. E. \& Hofmann, J. A.}{2008}]{King08} King, A. R.; Pringle, J. E. \& Hofmann, J. A., 2008, MNRAS, 385, 1621.

\bibitem[\protect\citeauthoryear{Reynolds, C. S.}{2013}]{Reynolds13} Reynolds, C. S., 2013, Classical and Quantum Gravity, 30, 244004.

\bibitem[\protect\citeauthoryear{S¸ adowski, A.; Narayan, R.; Penna, R. \& Zhu, Y.}{2013}]{sadowski13} S¸ adowski, A.; Narayan, R.; Penna, R. \& Zhu, Y., 2013, MNRAS, 436, 3856.

\bibitem[\protect\citeauthoryear{Volonteri, M.; Madau, P.; Quataert, E. \& Rees, M. J.}{2005}]{volonteri05} Volonteri, M.; Madau, P.; Quataert, E. \& Rees, M. J., 2005, APJ, 620, 69.

\bibitem[\protect\citeauthoryear{Younes, G.; Porquet, D.; Sabra, B.; Reeves, J. N. \& Grosso, N.}{2012}]{younes12} Younes, G.; Porquet, D.; Sabra, B.; Reeves, J. N. \& Grosso, N., 2012, AAP, 539, A104.

\bibitem[\protect\citeauthoryear{Yuan, F.; Gan, Z.; Narayan, R.; Sadowski, A.; Bu, D. \& Bai, X.-N.}{2015}]{yuan15} Yuan, F.; Gan, Z.; Narayan, R.; Sadowski, A.; Bu, D.\& Bai, X.-N., 2015, APJ, 804, 101.

\end{thebibliography}




\appendix




\bsp	
\label{lastpage}
\end{document}